# Photo-effect on ion transport in mixed cation and halide perovskites and implications for photo de-mixing

Gee Yeong Kim, Alessandro Senocrate, Yaru Wang, Davide Moia and Joachim Maier[*]

*Max Planck Institute for Solid State Research, Heisenbergstr. 1, 70569, Stuttgart, Germany*

[*]E-mail: office-maier@fkf.mpg.de

**Organic-inorganic hybrid perovskites are considered to be most promising photovoltaic materials. Highest efficiencies of perovskite solar cells have been achieved by using appropriate cation and anion mixtures. Mixed perovskite solar cells also show an improved stability. For both performance as well as stability, experimental information on electronic and ionic charge carriers is key, an information that so far has only been provided for methylammonium lead iodide; there we also found that light can enhance not only electronic but also ionic conductivities by more than one order of magnitude. We also proposed a mechanism for this surprising photo-ionic effect and explained its impact on photo-decomposition. Here we quantitatively deconvolute ionic and electronic transport properties for the practically relevant substitutions and mixtures. Specifically, we investigate various cation and anion substitutions (Cs; FA; Br) with a special eye on their photo-ionic effect. The results are not only of importance for light-induced degradation but also for light-induced demixing. As far as the photo-ionic effect is concerned, we find that the choice of the halide is of crucial importance, while the cationic substitutions are less relevant. The huge ionic conductivity enhancement found for iodide perovskites, is weakened by bromide substitution and eventually becomes insignificant for the pure bromide. Based on these experimental results, we provide a rationale for the experimentally observed photo-demixing.**

Methylammonium lead iodide (MAPbI$_3$) is the archetype of a whole class of halide perovskites that are in the focus of current photovoltaic research. High conversion efficiency cells are based on mixed cation and halide perovskites (*e.g.,* (FA,MA)PbI$_3$ and (Cs,FA,MA)Pb(I,Br)$_3$)[1,2] which show optimized combinations of stability and performances. In spite of the practical relevance of these mixtures and the impact of the transport properties on performance and stability, their transport properties are not systematically investigated. This especially holds for the ion transport.

For MAPbI$_3$, the various conductivity contributions have been precisely measured[3] and the nature of ion conduction could be safely attributed to iodide vacancies[4]. Contributions by methylammonium (MA)[4-6] and Pb ions are found to be much below the anion transport values. Of particular interest is the surprising finding that light illumination with energies above the bandgap can increase the ion conductivity –along with the electronic conductivity– by one or two orders of magnitude[7]. The photo-induced ion conductivity is assumed to be the consequence of hole localization that neutralizes iodide; the so-formed neutral iodine atoms can occupy interstitial sites where they are further stabilized by the polarizable environment in MAPbI$_3$[7-10]. As a consequence the concentration of iodide vacancies (i.e. of the ionic charge carriers) is largely enhanced and the ion conductivity thus increased. A detailed understanding of the photo effect on ionic transport in mixed perovskites is essential for stability and device performance.

In this contribution, we extract and separate ionic and electronic conductivities from d.c. polarization measurements performed on hybrid perovskite thin films in which the A-site cation or the anion is partially or fully substituted. Specifically we study MAPbBr$_3$, CsPbI$_3$, CsPbBr$_3$, as well as FAPbI$_3$ -MAPbI$_3$ and MAPbBr$_3$- MAPbI$_3$ mixtures.

As far as the ionic transport properties are concerned, one expects major differences if the iodine is replaced by the smaller and less polarizable bromine, but lesser variations if the A-site cation is substituted. As the polarizability is of major influence for the self-trapping of electronic carriers, major differences are expected to occur with respect to the photo-ionic effect, i.e. the light-induced ion conductivity. It is tempting -and we will come back to this point at the end of the paper- to connect this phenomenon with another highly debated effect, occurring in Br/I mixtures, viz. photo-demixing: When illuminated by light with above-bandgap energies, MAPb(I,Br)$_3$ mixtures segregate into I-rich and Br-rich domains[11-13].

As far as the structural and electronic properties are concerned, the various endmembers are expected to show far-reaching structural and electronic similarities but also some noteworthy differences. Notwithstanding hybridization details, in MAPbI$_3$ the valence band is predominantly formed by I-orbitals and the conduction band by Pb-orbitals. Given the same structure, A-site substitution then should not significantly change the band gap, rather result in geometric changes. In general, FAPbI$_3$ shows a structural instability at room temperature as it can crystallize either in a photo-inactive, non-perovskite δ-phase or a photoactive perovskite α-phase. In the purely inorganic CsPbI3 inversion to a photo-inactive orthorhombic phase is observed.[14] In the case of anion substitution, more significant changes of the electronic and ionic properties are expected. The replacement of I by Br perceptibly increases the bandgap from 1.5 eV to 2.3 eV (lower tendency of the bromide ion to get oxidized, *i.e.* lower valence band maximum)[15,16]. Under thermal equilibrium growth conditions MAPbBr3 shows p-type conductivity according to Kelvin probe force microscopy (KPFM) measurement and DFT calculation[15,17], but it is also expected to change to n-type at very low $Br_2$ partial pressures.

In order to separate and measure ionic and electronic conductivities in various perovskite materials, we performed systematic studies by using d.c. galvanostatic polarization experiments with ion-blocking electrodes and together with a.c. impedance experiments. If bulk polarization determines the process, which is always the case for sufficiently large samples, the steady state voltage is (apart from additional interfacial contributions) characterized by the electronic conductivity, while the initial voltage jump is determined by the total electronic and ionic conductivity. In selected cases we also cross-checked the ratio of ionic to electronic contributions by emf measurement.

Let us start with A-site substitution. In the dark, both FAPbI3 and CsPbI3 show mixed conduction similar to MAPbI3 (see **Figs. 1(a)** and **(c)**). The electronic conductivities of CsPbI3 and FAPbI3 are $2 \times 10^{-10}$ and $1 \times 10^{-9}$ S/cm, respectively, and of similar magnitude as for MAPbI3 (which for the same conditions shows a value of $1 \times 10^{-9}$ S/cm). Even though this follows the trend of the bandgap, it must be considered that electronic conductivity is primarily controlled by stoichiometry and doping content. According to previous work with MAPbI3 and defect thermodynamics, we thus rather expect, besides doping level and halide partial pressure, the formation energy of anion vacancies together with the valence band maximum to be decisive[3,18].

The ionic conductivities of CsPbI3 and FAPbI3 are $1 \times 10^{-10}$ and $2 \times 10^{-9}$ S/cm, respectively. Under the same conditions MAPbI3 shows a high value of $1 \times 10^{-8}$ S/cm and lower activation energy (0.4 eV) [7,19] than FAPbI3 (~0.6 eV, see SI). This trend is expected as the migration threshold should decrease with smaller size of the A-cation as it was observed for oxide perovskites[20]. An extreme case was recently reported for replacing MA with the large guanidinium cation resulting in a calculated migration barrier of 0.78 eV[22]. Also here it must be noted that measured activation energies contain a formation term which only disappears if the ion concentration is extrinsically dominated and association effects can be neglected.

Next we investigate the light effect on ionic and electronic conductivities for the A-cation substituted compositions. In all cases iodine is the anion, and light drastically increases both conductivities as shown in **Fig. 1(e)**. This similarity in the series of the iodide perovskites is expected in view of the fact that the phenomenon of ion conductivity enhancement is rather decoupled from the nature of the A-cation and primarily an issue of the anion. The transients of the d.c. polarization yields the chemical diffusion coefficient. We do not intend to discuss the time dependencies here in more detail, but it should be stated that the relaxation times are for all the materials not significantly varied by illumination as long as perovskites are composed of iodide anions. This may appear surprising but it is – as discussed for MAPbI$_3$ in Ref.[7] – rather very characteristic for a chemical diffusion, as the respective relaxation time is the product of a chemical resistance and a counteracting chemical capacitance. In the supplementary information this is discussed in greater detail.

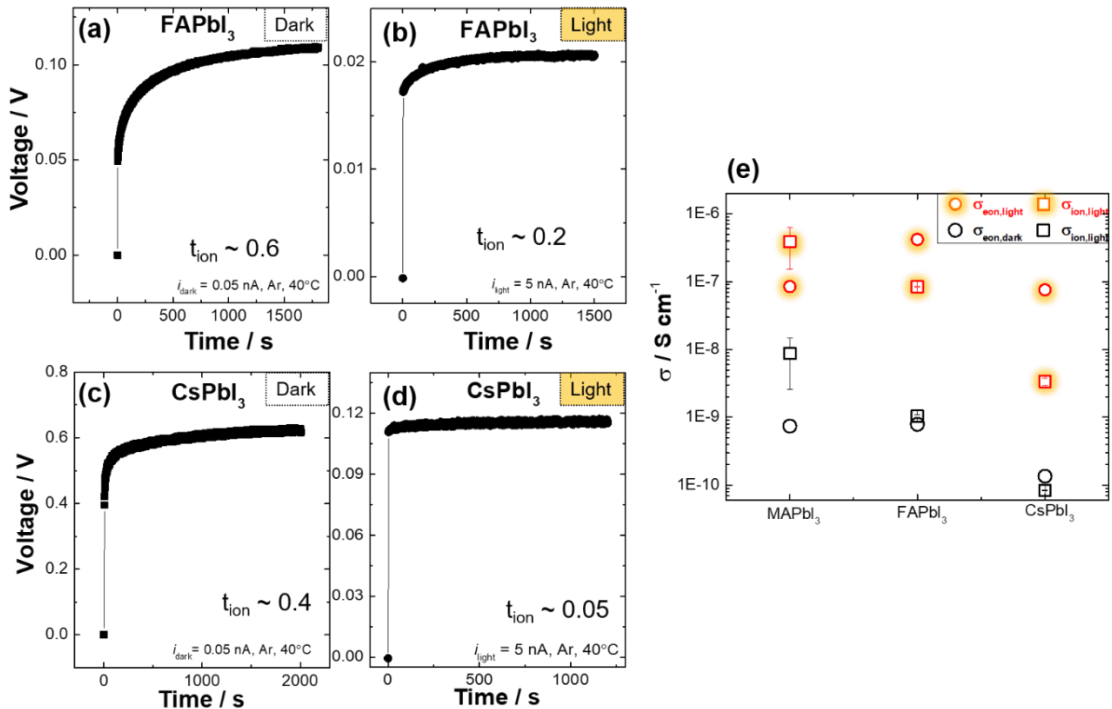

**Figure 1**. **Conductivity measurement on iodide perovskites.** (a)-(d) Stoichiometry polarization curve of FAPbI$_3$ and CsPbI$_3$ at 40°C under Ar atmosphere in the dark and under light (1 mW/cm$^2$).

(e) Conductivity variations of iodide perovskites. All these compositions show enhancement of both electronic and ionic conductivity under light. With respect to the error bars, see SI section 2.

As expected the conductivity measurements do not offer surprises for FAPbI$_3$/MAPbI$_3$ mixtures but exhibit a rather monotonic behavior as displayed in **Fig. 2**. The same is true for the relaxation times as we discuss in the supplementary information. An increased FA concentration ($x > 0.2$) leads to a lowering of the ionic conductivity under Ar. This is similar if data are considered under fixed iodine partial pressure (see supplementary). Comparable results had been obtained in an earlier preliminary study[3]. For FA- and Cs- compounds, illumination above the bandgaps leads to an enhancement of both electronic and ionic conductivity analogous to MAPbI$_3$, as shown in **Fig. 2(a)** and **(c)**. This similarity also holds for the light intensity dependence.

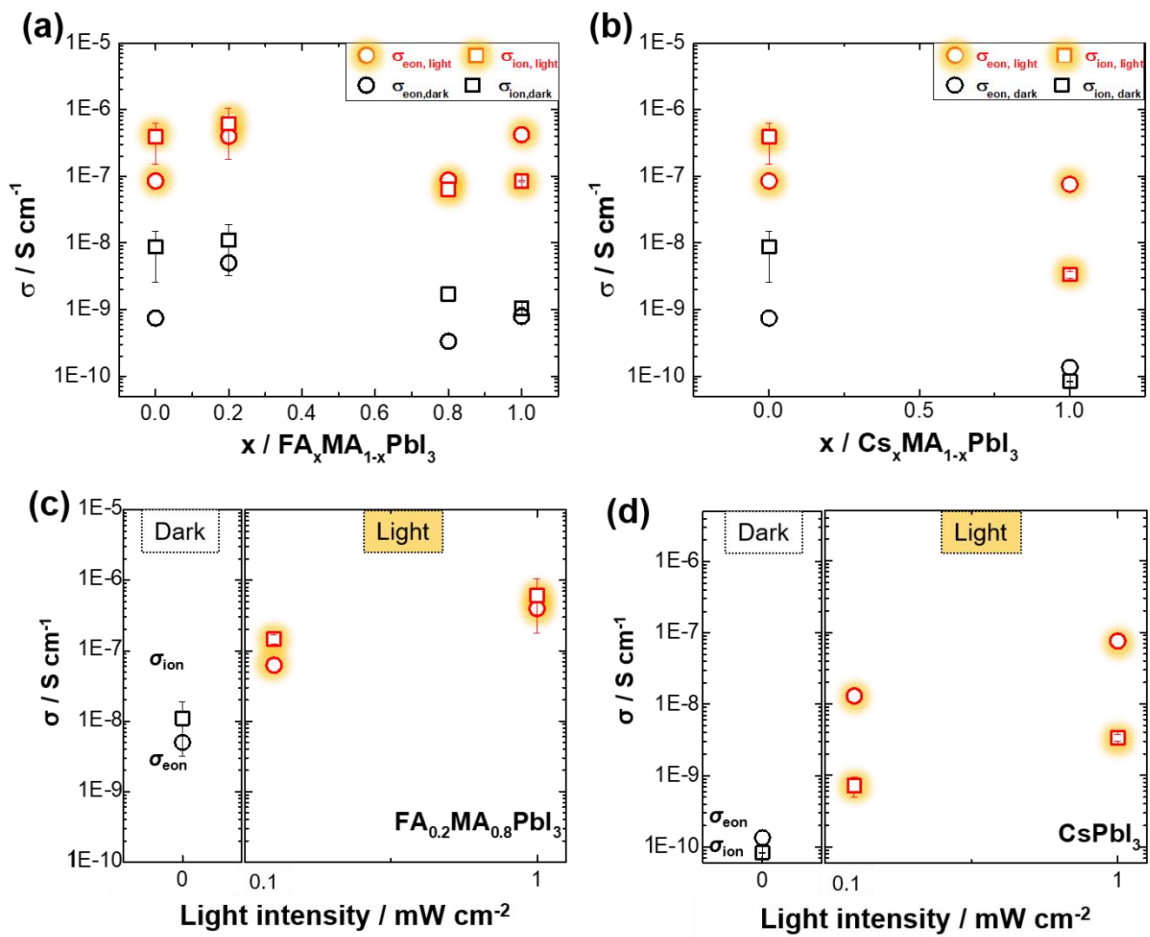

**Figure 2. Electronic and ionic conductivity of perovskite cation mixtures in the dark and under light.** (a) (FA$_x$MA$_{1-x}$)PbI$_3$ and (b) CsPbI$_3$. In (a) and (b), we use a Xenon lamp for illumination with 1 mW/cm² of light intensity. (c) and (d) indicate both electronic and ionic conductivity of iodide based perovskites as a function of light intensity. With respect to the error bars, see SI section 2.

Now we turn to the replacement of iodine by bromine and the respective mixtures (**Fig. 3**). Again the variation of the electronic conductivity is parallel to the bandgap variation, and presumably (and more importantly) to the variation of the valence band maximum (extrinsic hole conduction). The activation energy for MAPbBr$_3$ (measurement under Ar atmosphere) is smaller by ~0.1 eV than for MAPI (~0.4 eV)[7], reflecting the lower migration threshold for the Br-vacancy motion owing to smaller size. This is in line with results obtained for PbI$_2$ and PbBr$_2$[19]. Like in the binary halides, it appears that in halide perovskites there is some compensation between mobility and vacancy concentration resulting in similar conductivity values. A sharp contrast to MAPbI$_3$ is observed for the light effect on MAPbBr$_3$. While the electronic conductivity is enhanced by more than one order of magnitude in MAPbBr$_3$ illuminated with 1 mW/cm², the effect on the ionic conductivity is weak if not zero within the error margins. As a consequence the ionic transference number under light is 50-90%. This much smaller self-trapping may also be the reason for the much quicker polarization (see **Fig. 3(b)**, as discussed in supplementary). This smaller relaxation time corresponds to a larger chemical diffusion coefficient of the halogen and predicts a quicker out-diffusion of the halogen. This point is taken up again at the end of the paper.

At this point it should be emphasized that the analysis of the d.c. experiments in this work assumes the observed changes in voltage upon application of a current step to halide perovskite films between gold contacts to be related to ideal stoichiometric polarization in the perovskite mixed conductor. It therefore assumes that no change in electronic conductivity is occurring on ionic redistribution. We note that variation in electron hole recombination and injection processes due to ion redistribution in the sample could play a role in the changes of the measured voltage under light, as it explains many of the features of halide perovskites when used in solar cell devices[23-25]. The contribution of this effect for our measurement is currently under investigation and might not be significant when considering the symmetric samples presented here.

To double-check that our measurements under light and in the dark indeed yield correct ionic and electronic conductivities, we performed *emf* measurements (transference number measurements). The chemical gradient (chemical potential of bromine)[3,7,26] was established by using two different mixtures in pellet form, Cu/CuBr and Pb/PbBr$_2$, as electrodes. The open circuit voltage of this "battery cell" was measured under dark and light conditions. The results corroborate the presence of significant ion conductivity in the dark and also under light. Note again that the open circuit voltage would be zero or on the level of possible thermo- or photo- voltages, if the conductivity was purely electronic. Nonetheless the exact deconvolution of *emf* and possible photovoltage contributions to the measured voltage deserves future investigations.

As shown in **Fig. 3(c)**, we derive a $t_{ion}$ value of ~0.8 for the ionic transference number ($t_{ion}=\sigma_{ion}/(\sigma_{eon}+\sigma_{ion})$) in the dark and of ~0.4 under light when we refer to similar partial pressure as in the polarization experiment. Given the fact that these values have to be somewhat corrected to higher values and that the conditions are not exactly the same, the agreement with the polarization results is satisfactory. Details of this evaluation which is complicated by the fact that very low partial pressures are established at the contacts but the initial high partial pressure value is still expected to dominate in the interior, are given in the supplementary.

The UV-Vis/toluene experiments[7,27] that we used in the iodine case for revealing the kinetics of stoichiometric change were not applicable here owing to the insufficient Br amount in the solvent (see supplementary). We attribute this primarily to the lower homogeneity range of MAPbBr$_3$ (smaller achievable anion deficiency), also reflected by the fact that the chemical capacitance of MAPbBr$_3$ is one order of magnitude smaller than MAPbI$_3$. The photo-effect on CsPbBr$_3$ is rather analogous to MAPbBr$_3$ as shown in **Fig. 3(f).** The observed insensitivity of the ionic conductivity on illumination for the bromide, which is in striking contrast to the iodide, is consistent with our enhancement mechanism that involves the anion directly. Bromine is less polarizable and thus a stabilization of neutral interstitial defects is less likely.

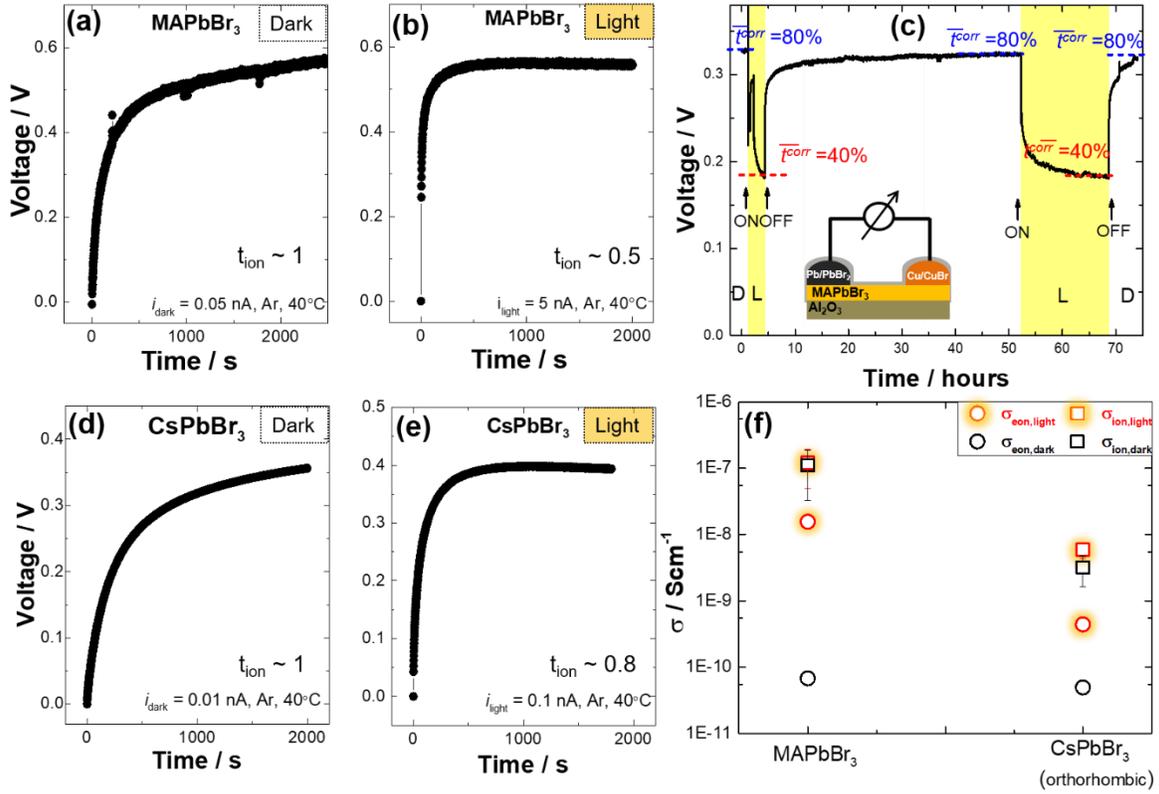

**Figure 3. D.C. galvanostatic polarization measurement with ion blocking electrode at 40°C under Ar atmosphere in the dark and under 1 mW/cm² of light.** (a), (b), (d), (e) polarization results of MAPbBr₃ and CsPbBr. (c) open circuit voltage measurement (transference measurement) on Cu,CuBr/MAPbBr₃/Pb,PbBr₂ cell in the dark and under light at 60 °C. (f) shows the conductivity variations of MAPbBr₃ and CsPbBr₃ under light as extracted from polarization measurement. With respect to the error bars, see SI section 2.

We now refer to the conductivity measurements on anion mixtures. It was reported that replacement of I with 20% of Br induces a phase change from a tetragonal to cubic phase[28]. In the anion mixtures (MAPb(I₁₋ₓBrₓ)₃) the band gap can be tuned from 1.6 eV to 2.4 eV by controlling $x$[15,28]. The electronic conductivity is monotonically decreased on increasing Br concentration; again this is parallel to the bandgap but might rather be connected with the course of the maximum of the valence band which directly influences the hole formation energy. The ionic conductivity is monotonically increased on increasing the Br content, as expected from the high values of MAPbBr₃ and the absence of any crystallographic anomaly.

We had already mentioned the striking difference to MAPbI$_3$ occurring under illumination. The results for the ionic conductivities under illumination in anion mixtures is shown in **Fig. 4(a)**, demonstrating that the ionic conductivity enhancement by light is gradually reduced with increased Br content. Pure MAPbBr$_3$ displays only a weak increase of the ionic conductivity (not more than a doubling), while the electronic conductivity increases by two orders of magnitude under light with the same light intensity (the wavelength dependence is given in the supplementary, Fig. S5). Zhou *et al.* reported activation energy results on CsPbI$_2$Br and concluded that the effect of light-induced ion migration can be reduced by inorganic Cs substitution[21]. Irrespective of the fact that the activation energies only reflect the temperature dependence of the conductivity, our results clearly show that Cs-substitution does not significantly impact the light effect, but Br-substitution does.

Considering the course of the conductivities (**Fig. 4(a)**) as well as the transients of the polarization curves (see **Fig. 5**) within I/Br mixtures one realizes two anomalies:

(i) Strikingly, the nominal I$_{50}$Br$_{50}$ mixture shows a clear minimum in the electronic and ionic conductivity under light.

(ii) The observed stabilization time is anomalously large (in fact the data point in **Fig. 5** refers to a lower bound).

Both findings, that are absent under dark conditions, are in agreement with the photo-induced de-mixing into I-rich and Br-rich phases reported in literature[11,13,29-34]. In such a case, long-time morphological changes occur which are connected with geometrical blocking effects. Then the time dependence does not follow a polarization behavior and does not allow for an extraction of a defined relaxation time and hence chemical diffusion coefficient (unlike for end members and mixtures in the dark). These issues, and in particular the appropriateness of the stabilization time to map out the phase width, will be subject of ongoing work where in particular the two time scales (polarization and de-mixing) have to be separated.

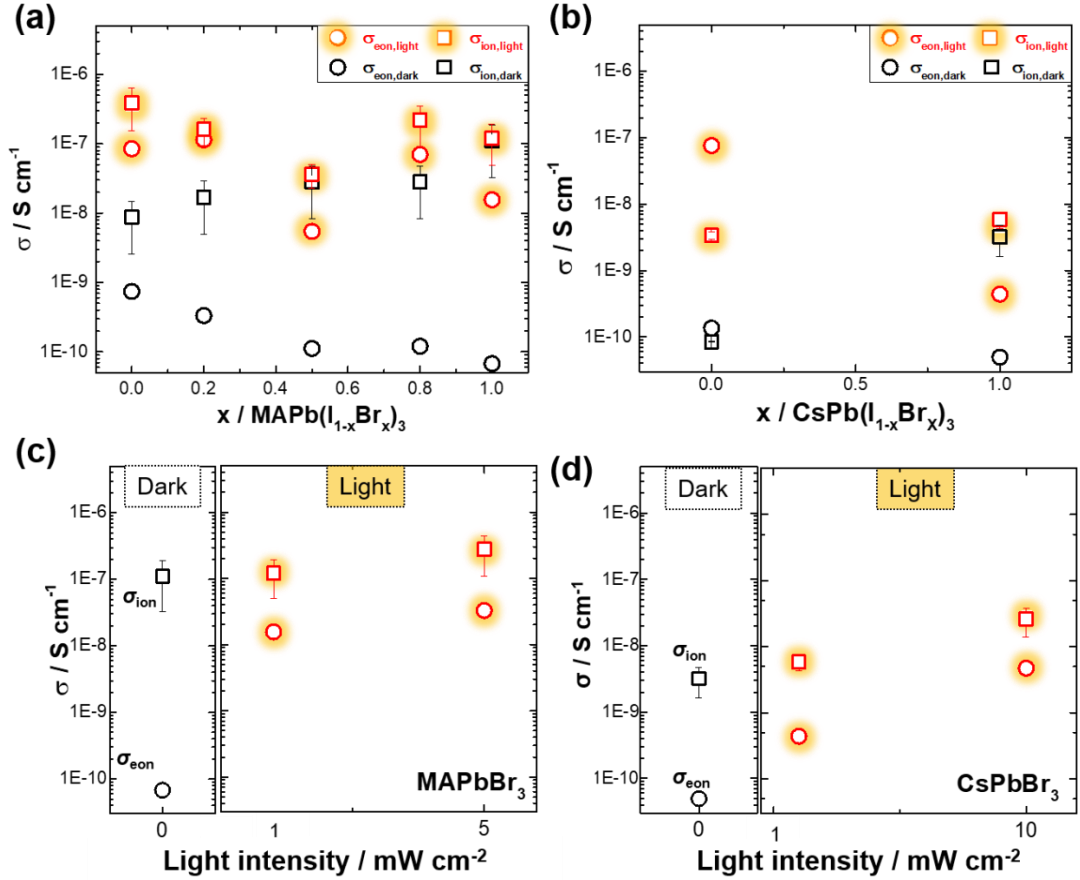

**Figure 4**. **Electronic and ionic conductivity of perovskite anion mixtures in the dark and under light.** (a) MAPb($I_{1-x}Br_x$)$_3$ and (b) CsPb(I,Br)$_3$ (with light intensity of 1 mW/cm$^2$). Comparison of conductivities as a function of light intensity in (c) MAPbBr$_3$ and (d) CsPbBr$_3$ under Ar atmosphere at 40°C. Data are extracted from *d.c.* polarization measurement. In terms of error bars estimation, see SI section 2.

In order to explain the de-mixing effect that is well reported in the literature, several explanations have been proposed. DFT calculations suggested[12] a thermodynamic phase instability at room temperature in the dark; accordingly the light effect consists in enhancing the kinetics of de-mixing. The enabled de-mixing under light should then however not be reversible at all. In contrast the experimentally observed –at least partial– reversibility indicates a thermodynamic effect; *i.e.* a miscibility gap opens –or at least widens– under illumination. This partial reversibility is also found in our work. One thermodynamic explanation was based on the band-gap variation[12]. Since we cannot simply assume that the excess energy of the illumination over the band gap excitation remains stored inside, this argument does not seem to generally apply. Other explanations refer to strain effects[34,35], and explain the demixing by a lowering of the mechanical energy.

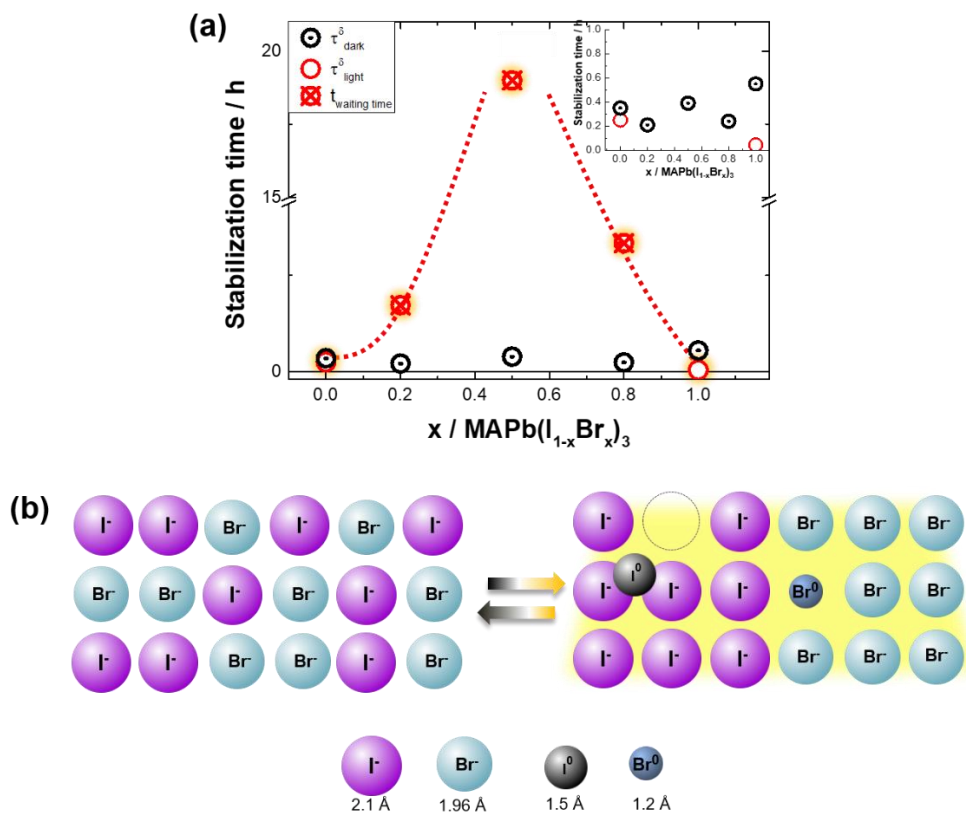

**Figure 5.** (a) Stabilization times of anion mixtures in the dark (open symbols) and under light illumination (closed symbols with hollow). In the case of the endmembers and also generally in the dark the evaluation of the stoichiometric polarization yields a well-defined relaxation time that can be converted into a chemical diffusion constant. The values for the mixtures under light represent simply stabilization times of conductivity measurement (1/e criterion). The middle value even only represents a lower limit. This clear anomaly when compared to the end-members or to the dark situation can be attributed to demixing. Only the time constant (diffusion coefficient) for the pure bromide under light is significantly smaller (larger) (cf. supplementary) (b) The schematic image shows that the occurrence of hole self-trapping in an iodine environment favors (reversible) photo-demixing in mixture. The situation in the right part is simplified as the interstitial neutral iodine is further stabilized by ionic rearrangement (dumbbells *etc.*).

From the suggested mechanism of photo-enhanced ion conduction, and from the difference observed for iodide and bromide perovskites, it appears natural to invoke the distinct chemical effects (*i.e.* self-trapping of holes as proposed in refs[7,8] in the iodide but not in the bromide) for an understanding of the photo-demixing. It is relevant here to briefly summarize our present picture of the photo-ionic effect.

We assume that in the case of the iodide the hole is localized (loss of delocalization energy) and then a self-trapped hole defect is formed that is stabilized by electronic but in particular by ionic polarization in the form of dumbbell structures or higher order aggregates[8,9]. Owing to a much lower polarizability and in line with recent calculations[8], the effect is much lower and probably absent in the bromides. (In view of the high ionic conductivity contribution in $MAPbI_3$ under light, it is highly likely that also excess electrons are trapped presumably at Pb-sites (see also [36,37])).

As the incorporation of these defects introduces a negative Gibbs energy contribution ($\Delta G^0 + \Delta G_{\text{configuration}}$), and only becomes effective if I-rich and then also Br–rich clusters are present or formed, the segregation is favored. This is valid irrespective of $\Delta G^0$, the non-configurational part, being positive or negative. The assessment of the magnitude of the effects suggest $\Delta G^0$ being negative and the number of self-trapped holes being very large. Probably the number of trapped conduction electrons is large as well. This seems to be supported by the recent finding of a partially reversible appearance of neutral iodine and neutral lead on illumination[11,12]. A regular solution model (see supplementary) indicates that the effective energy gain is on the order of several tens of meV, which suffices to destabilize the mixtures. This would also explain the disappearance of the de-mixing in grains of nanometer size[38-40], as then the energy of forming an interface would outweigh the bulk free energy gain due to de-mixing (see analogous situation in $LiFePO_4$[41]). More details are found in the supplementary. An in-depth discussion of the mixture effects including strain effects must be reserved for a forthcoming paper.

At the end, let us comment on the observed higher stability of the bromide in the context of these transport studies. As addressed above and detailed in the supplementary, the larger trapping in the iodide case appears to be the main reason for the lower chemical diffusivity of iodine when compared to bromine. On the other hand it can lead to a lower probability of recombination and thus to a higher increase of the halogen chemical potential. If we only refer to the diffusion kinetics of the intrinsic photo-degradation, the bromide should be even less stable. At room temperature, however, the surface reaction step typically plays a decisive role[42,43]. Whether it is a lower surface reaction rate constant that provides the higher stability of the bromide or whether it is a matter of a lower driving force is under investigation. It is worth emphasizing that there are, in addition to the intrinsic degradation, various extrinsic channels of importance, e.g. the interaction with oxygen or water traces; in these cases significant differences in the driving force exist between iodide and bromide[44].

**Conclusions**

In conclusion, we performed a systematic study to investigate the electronic and ion conduction under light in various pure and mixed lead halide perovskites of theoretical and practical interest. We considered A-site substitution of MAPbI$_3$ by FA and Cs but also of the anion by Br. Special emphasis was laid on the photo-effect on the ion conduction. The A-site substitution was of only small influence, while anionic substitution was of great influence and showed a significant lowering of the photo-effect on ion transport. This not only corroborates our electron-ion coupling mechanism for the photo-induced ion transport, it also provides a rationale for a thermochemical route to the photo de-mixing phenomenon of halide perovskite mixtures by the fact that the self-trapping free energy can only be capitalized if iodide rich domains (and hence bromide-rich domains) are formed. (The energetic effects are rather small and easily overcompensated by interfacial costs resulting in a vanishing miscibility gap for a nano-sized crystal.)

## Author Contributions

G.Y.K. was responsible for sample preparations and performed characterizations. G. Y. K. and Y. W. performed conductivity measurements. J.M. supervised the work. G.Y.K., A.S., Y. W., D. M. and J.M. discussed the results and wrote the manuscript.

## Competing interests

The authors declare no competing financial interests.

## Data availability

The data that support the findings of this study are available from the corresponding author upon reasonable request.

## Methods

*Preparation of perovskite thin-films for electrical measurements*

MAI and MABr were obtained using the reported procedure by Im *et al.*[45] An equimolar solution (1 M) of MAI, MABr, FAI, CsI, CsBr (Sigma Aldrich) and $PbI_2$, and $PbBr_2$ (Alfa Aesar) in DMSO was prepared and spin-coated on polished sapphire (0001) substrates, previously equipped with interdigitated Au electrodes (2 µm spacing, 100 µm length). Prior to the spin-coating step, substrates were cleaned with an $O_2$-plasma treatment, in order to remove organic impurities and enhance the hydrophilicity of the surface. Perovskite thin films were coated by spin-coating process at 65 rps and 150 rps for 2s and 180s on $Al_2O_3$ substrates, respectively[45,46]. During spin-coating, a drop of chlorobenzene was used to induce rapid crystallization,[46] and the sample was later annealed for 3 minutes at 100 °C. In case of $CsPbI_3$, we annealed the sample for 30 min at 350 °C. The thickness of films is around 200 nm. All procedures were carried out in an Ar-filled glovebox ($O_2 < 0.1$ and $H_2O < 0.1$ ppm).

*D.c.-galvanostatic polarization*

*D.c.* polarization experiments were performed by using a source meter (Keithley model 2634b) and by monitoring the voltage change. Measurements were carried out in the dark and under light illumination and by accurately controlling temperature and atmosphere over the sample (oxygen content and humidity were monitored using appropriate sensors). We varied the composition by controlling the nominal concentration of the precursors in the solution and measured the voltage via Au ion-blocking electrodes in an interdigitated arrangement. Au electrodes were considered as ideally electronically selective electrodes. As the light source, a Xenon arc lamp was used (~1 mW/cm$^2$), and its intensity was calibrated using a power meter (Thorlabs).

*Iodine partial pressure dependence*

To control $P(I_2)$ over the samples, argon was used in a container with solid iodine chips, kept in a thermostat at a fixed temperature (always below room temperature, between $-$ 35°C and -5°C).

The iodine partial pressure was assumed to correspond to the equilibrium pressure of iodine at the thermostat temperature, which was calculated based on a published equation[47]. Similar values are obtained by estimating $P(I_2)$ purely from thermodynamic considerations, starting from Gibbs free energy of sublimation of solid iodine.

*EMF measurement*

A MAPbBr$_3$ thin film was contacted with pellets, one that consisted of a physical mixture of Cu and CuBr and the other of Pb and PbBr$_2$. The cell was kept under an Ar atmosphere at 60℃. The two electrodes were connected to a high impedance source meter (Keithley model 2634b). The cell was sealed using PMMA to avoid any exchange with the external atmosphere. The light source is a LED lamp with 420 nm of wavelength and 4 mW/cm$^2$ of intensity.

*XRD measurements*

We obtain data by using a PANalytical Empyrean Series 2 (Cu Kα radiation, 40 kV, 40 mA) in grazing incidence (ω=2°). X-ray diffraction patterns were acquired in an Ar filled beryllium dome.

*Toluene experiments*

MAPbBr$_3$ films are immersed in toluene in a quartz cuvette. The bromine amount lost by the sample is measured by in-situ optical absorbance as a function of time, using a *UV-Vis* spectrophotometer (Cary *UV–Vis* 4000 spectrometer). Illumination was achieved with a tungsten lamp that delivers approximately 1 mW/cm$^2$ light intensity over the samples.

[Supplementary Materials]

# Photo-effect on ion transport in mixed cation and halide perovskites and implications for photo de-mixing

## **Table of contents**

# 1. Material properties

Figure S1 shows XRD patterns of mixed cation and anion perovskite thin films on $Al_2O_3$ substrates. Anion mixtures indicate no secondary phase such as $PbI_2$, however we observe small amount of $PbI_2$ in $I_{20}Br_{80}$ sample. Cation mixtures (FA,MA) show no secondary phase, but $FAPbI_3$ has a certain amount of delta phase and $PbI_2$. This is due to structural instability at low temperature (delta phase is favorable at low T). For the $CsPbI_3$, the sample was cubic-phase with black color; however the phase partially changed to orthorhombic during XRD measurement. At least, $CsPbI_3$ cubic-phase is stable under Ar atmosphere ($O_2 < 20$ ppm) for electrical measurements.

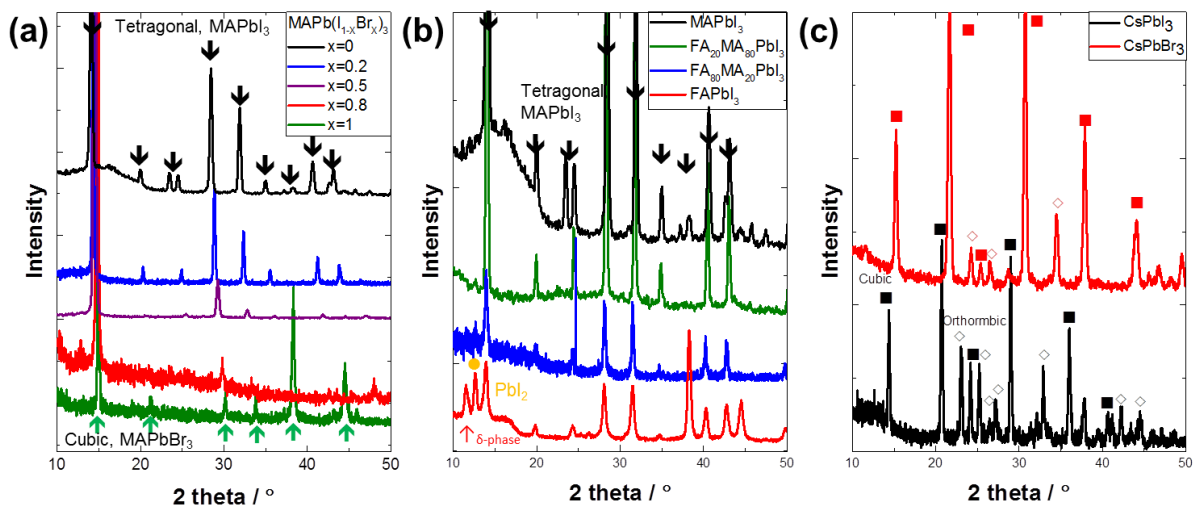

**Figure S1**. XRD patterns of (a) $MAPb(I_{1-x}Br_x)_3$ (b) $(MA_{1-x}FA_x)PbI_3$ (c) $CsPb(I,Br)_3$ thin films on $Al_2O_3$ substrates.

## 2. Analysis of *d.c.* polarization measurements

Here we show the consistency check from the comparison of transient and steady state based on *d.c.* polarization curve. The observed polarization curves in the dark and under light are shown in main text. The polarization transient is in the ideal case characterized by a short time square-root time-behavior ($\sqrt{t}$) and a long-time exponential time-behavior (characterized by a time constant that is inversely proportional to the chemical diffusion coefficient). Figures below are an example of a polarization curve and time constant fitting in the dark and under light and its fitting according to the above. As shown in Figs. S2-S3 (a), (c), (e), and (g), the curve indicates the square root t behavior for short time in the dark and under light and Figs. S2-S3 (b), (d), and (f) show long-time exponential behaviors. The time constant from the long time fitting values are in line with short time fitting values. (For more details on the technique the reader is referred to literatures[1-3] and references therein.)

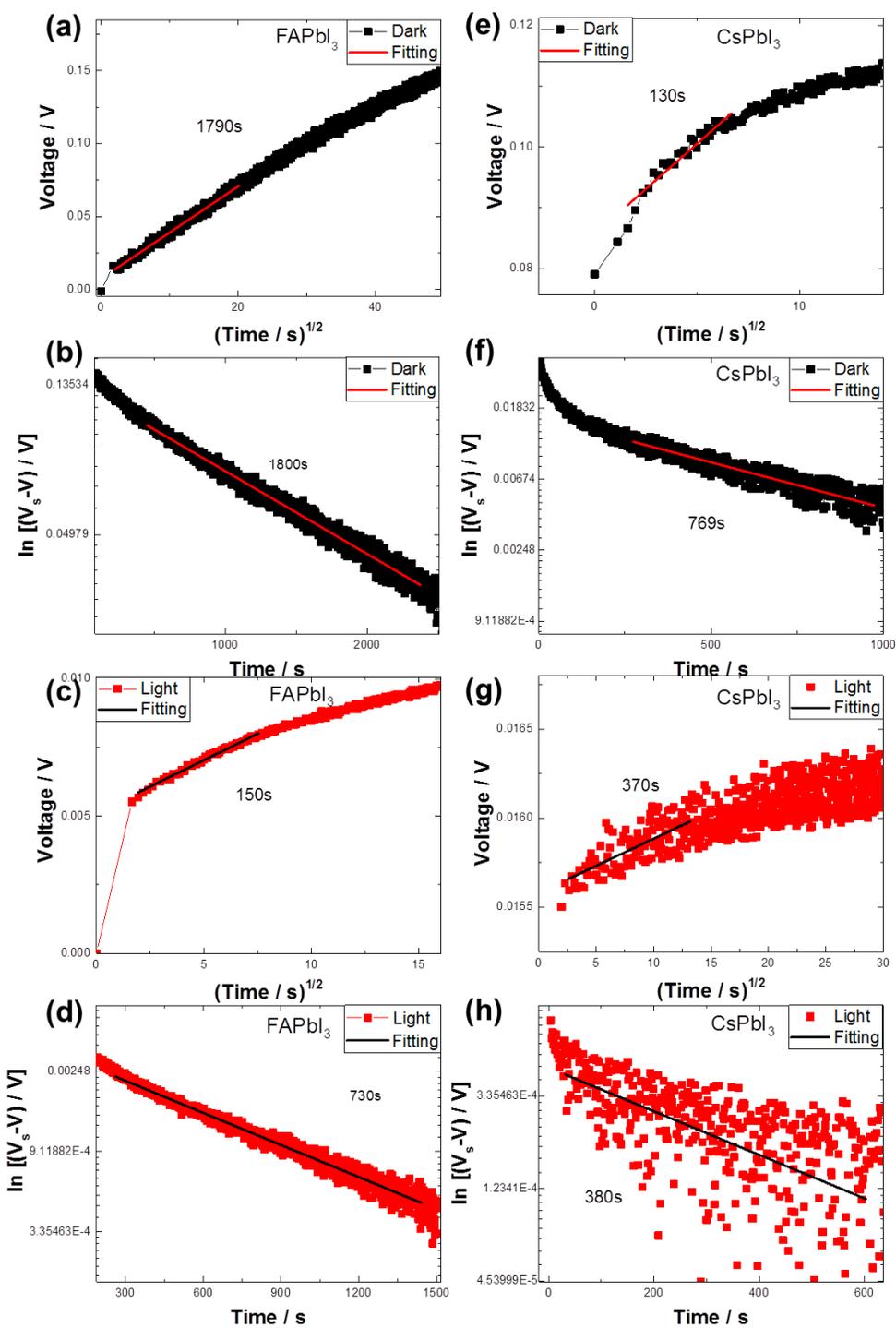

**Figure S2**. Stoichiometric polarization curve of FAPbI$_3$ and CsPbI$_3$ thin films in the dark and under light and fitting with short and long-time scale.

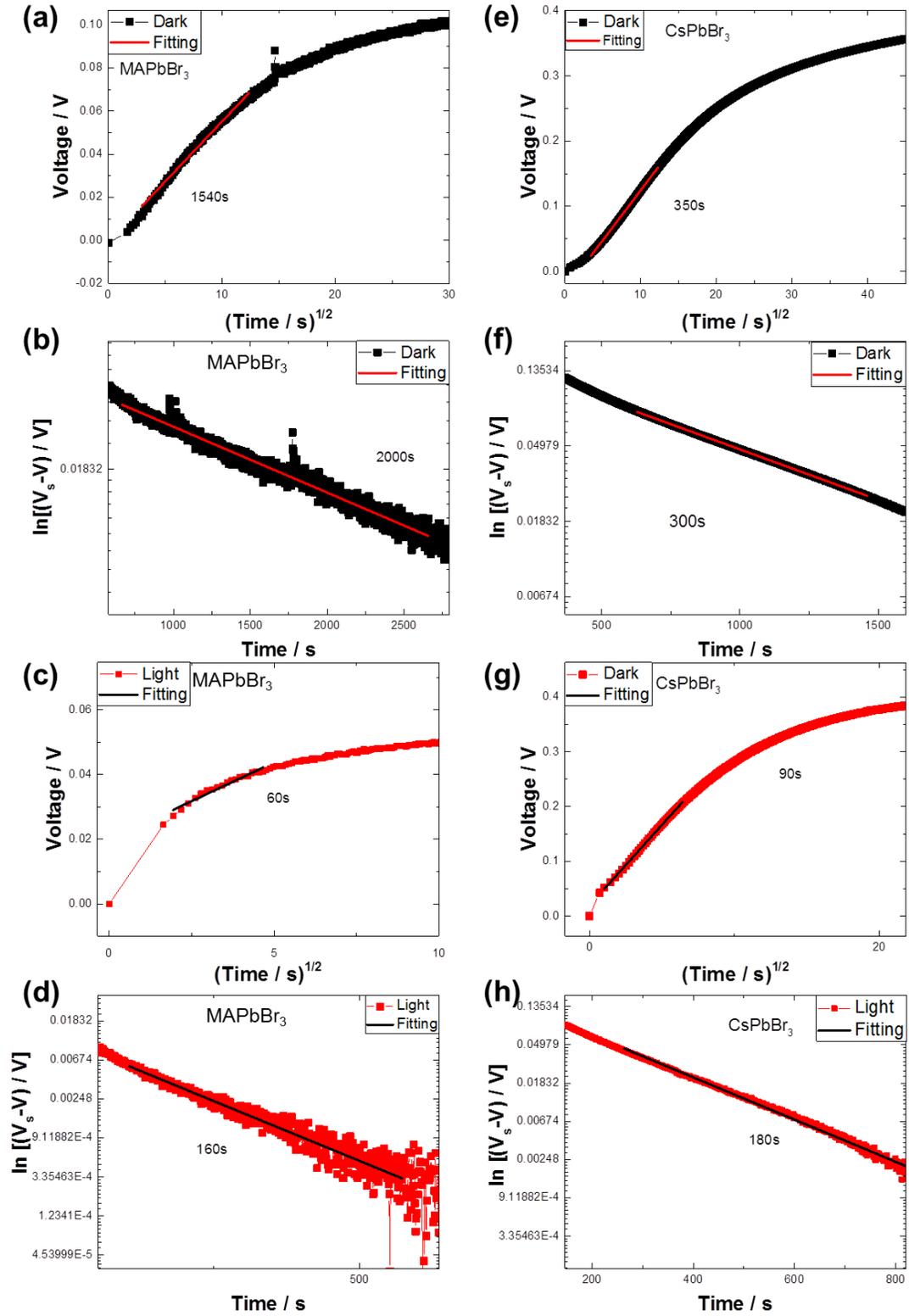

**Figure S3**. Stoichiometric polarization curve of Br-based perovskite thin films in the dark and under light and fitting with short and long-time scale.

In terms of dc polarization measurement, we have to address here the values of electronic conductivity (steady state voltage) is accurate, however, the evaluation of the ionic conductivity from the first jump of voltage can give significant errors in the evaluation of the ionic conductivity due to their limited time resolution when the sample has high ionic transference number. In order to resolve this issue, we fit the early time scale voltage response of dc polarization measurements to a function $V_{fit}(t) = A\sqrt{t} + V_0$ and extrapolate the value of $V_{fit}(t = 0) = V_0$ from the fit (see Figs. S2-S3). From the first data point measured upon application of the dc current $V(t = t_{res})$ ($t_{res}$ is the time resolution of the dc measurement) and $V_0$, we can evaluate the fractional error associated to the estimate of total conductivity as $\frac{\Delta V}{V} = \frac{\Delta\sigma_{tot}}{\sigma_{tot}} = \frac{V(t=t_{res})-V_0}{V(t=t_{res})}$, and from which we estimate the error on the ionic conductivity $\frac{\Delta\sigma_{ion}}{\sigma_{ion}}$. We find that this error is negligible for measurements with low $t_{ion}$ ($\frac{\Delta\sigma_{ion}}{\sigma_{ion}} < 0.6$ for $t_{ion}$<0.9) but can be significant for samples showing large $t_{ion}$. Therefore, in the main text, we report values of ionic conductivity that are calculated in the following ways:

i)     $\sigma_{ion}$ calculated directly from total resistance the dc polarization measurement for cases where $t_{ion} < 0.9$.

For samples where $t_{ion} > 0.9$, $V(t = t_{res})$ can be small and $V_{fit}(t = 0)$ is very close to 0 V. In this case, we measure the impedance spectrum of the sample and estimate the ionic conductivity as below.

ii)     $\sigma_{ion}$ was calculated using a value of $\sigma_{tot}$ extracted from the high frequency semicircle in the Nyquist plot.

### 3. Chemical capacitance and diffusion coefficient

The chemical diffusion coefficient ($D^\delta = \dfrac{L^2}{(\pi^2 \tau^\delta)}$) where L is thickness of sample and $\tau^\delta$ is time polarization time constant and chemical capacitance (resistance) ($C^\delta = \dfrac{\tau^\delta}{R^\delta}$) have been extracted from time constants as indicated in Figs. S2-3. What we found is a similar diffusion coefficient in MAPbI$_3$ and FAPbI$_3$ both in the dark and under light illumination (see Fig. S4a). However, the diffusion coefficient of MAPbBr$_3$ and CsPbBr$_3$ is increased by one order of magnitude by light illumination (see Fig. S4c). The comparison of the chemical diffusion coefficients of MAPbI$_3$ and MAPbBr$_3$ give the following situation: The values for MAPbI$_3$ or MAPbBr$_3$ are similar in the dark. The MAPbI$_3$ value is slightly higher under light ($\sim$ factor 2) but much higher for MAPbBr$_3$ under light ($\sim$ factor 10). In all cases the ionic conductivity is larger or comparable with the electronic one speaking for a higher ionic defect concentration. Self-trapping is substantial for MAPbI$_3$ but not for MAPbBr$_3$.

In the following we distinguish between two types of trapping. *Intrinsic self-trapping* $h^\bullet \rightleftharpoons h_{Tr}^\bullet$ as occurring in MAPI, for which we simply obtain

$$\chi_p = \frac{\partial [h^\bullet]}{\partial [h^\bullet]_{Tr}} = \frac{1}{K_{Tr}} = \frac{[h^\bullet]}{[h^\bullet]_{Tr}}$$

where $\chi_p$ is the differential (un) trapping factor and $K_{Tr}$ the trapping mass action constant.

Beyond that we have to be aware of *extrinsic trapping* which is then particularly important for MAPbBr$_3$. Unlike under light, where the electronic concentration in MAPbBr$_3$ should exceed the extrinsic trap concentration, we can assume that in the dark, an extrinsic trap controls the hole concentration according to $T' + h^\bullet \rightleftharpoons T^x$ where $[T'] + [T^x] = T = const.$ at equilibrium. $T'$ is the free traps and $T^x$ the occupied traps. Such a trap maybe e.g. oxygen $\left( O_{Br}^{'} \text{ or } O_{Br}^{\sim} \right)$.

The chemical diffusion coefficient $D^\delta$ can be written as[4,5]

$$D^\delta = t_v D_p \chi_p + t_p D_v \chi_v$$

$$\frac{F^2}{RT} D^\delta = t_v \sigma_p \left( \frac{\chi_p}{c_p} + \frac{\chi_v}{c_v} \right) = t_p \sigma_v \left( \frac{\chi_p}{c_p} + \frac{\chi_v}{c_v} \right)$$

where $t_v$ and $t_p$ are transference number ($t_v \equiv \frac{\sigma_v}{\sigma}, t_p \equiv \frac{\sigma_p}{\sigma}$), $\chi_v$ and $\chi_p$ denote the differential trapping factors of vacancy and hole carriers.

If $\left[ T^x \right] \gg \left[ h^{\cdot} \right]$, i.e. extrinsic traps dominate, this equation can also be written as[1]

$$\frac{F^2}{RT} D^{\delta} = t_v \sigma_p \left( \frac{1}{\left[ T^{\cdot} \right]} + \frac{1}{\left[ T^x \right]} + \frac{1}{[c_{V_I^{\cdot}}]} \right)$$

If we assume $\left[ T^x \right] \gg \left[ T^{\cdot} \right]$ and $\left[ T^{\cdot} \right] = \left[ V_I^{\cdot} \right]$ owing to electroneutrality, then $D^{\delta} = t_p D_v$.

Under light $\left[ h^{\cdot} \right] \gg T$ and $t_v$ is still high (0.5 for experimental result) then,

$$D^{\delta} = t_v \chi_p D_p$$

for MAPbBr$_3$ where self-trapping is small $\chi_p$ can be set to 1; $D^{\delta}$ then turns out to be indeed much higher than in the dark ($D_p \gg D_v$) and

$$D^{\delta} \approx 0.5 D_p.$$

$D_p$ and $D_v$ are the individual diffusivities of hole and vacancy. For MAPI, $\chi_p$ is rather small and $D^{\delta}$ greatly depressed with respect to $D_p$ explaining the lower $D^{\delta}$. As in MAPI self-trapping can be assumed to be dominant with respect to extrinsic trapping, we obtain the same result as under illumination ($\sim D_p / K_{T_r}$), in agreement with the experiments. The similar results for $D^{\delta}$ (MAPbI$_3$) at $D^{\delta}$ (MAPbBr$_3$) in the dark must be rather coincidental and the outcome of compensation between trapping and mobility. Along the same lines: The chemical capacitance in MAPbBr$_3$ is about one order smaller in the dark than MAPbI$_3$, while for the chemical resistance the situation is reversed.

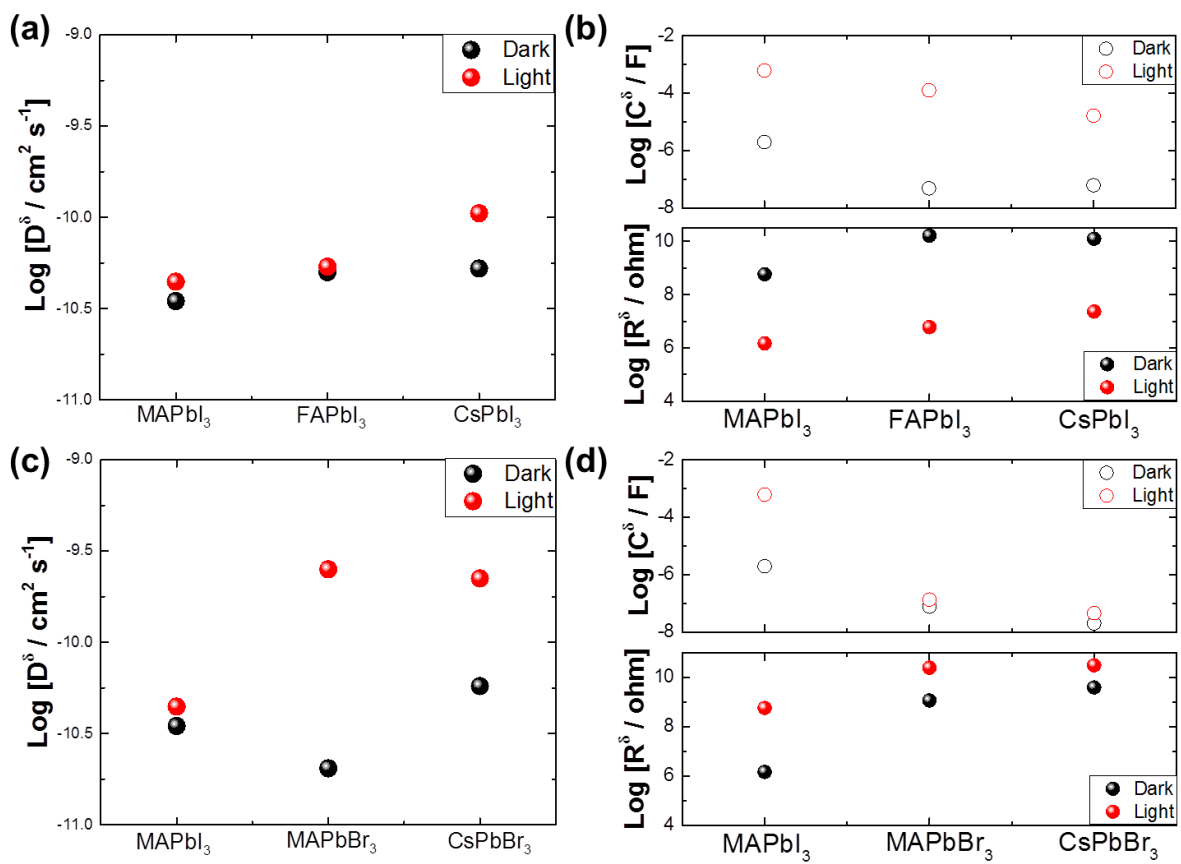

**Figure S4**. (a) Chemical diffusion coefficient ($D^{\delta}$) and (b) chemical capacitance ($C^{\delta}$) and resistance ($R^{\delta}$) of iodide perovskites as well as (c) $D^{\delta}$, (d) $C^{\delta}$ and $R^{\delta}$ of bromide perovskites in the dark and under light illumination.

## 4. Wavelength dependence conductivity measurements

In order to confirm the dependence of photon energy, we performed conductivity measurement with different wavelength of light. What we consistently found is that the effect on ion conduction only sets in, if the wavelength is smaller than the critical value set by the band gap. The yellow area indicates photon energy is higher than bandgap. If the photon energy is larger than the bandgap, the ionic conductivity enhancement is even larger than the white light in MAPbI$_3$. For MAPbBr$_3$, all the wavelength of light shows no enhancement of ionic conductivity, but only increase of electronic conductivity by 420 nm of light.

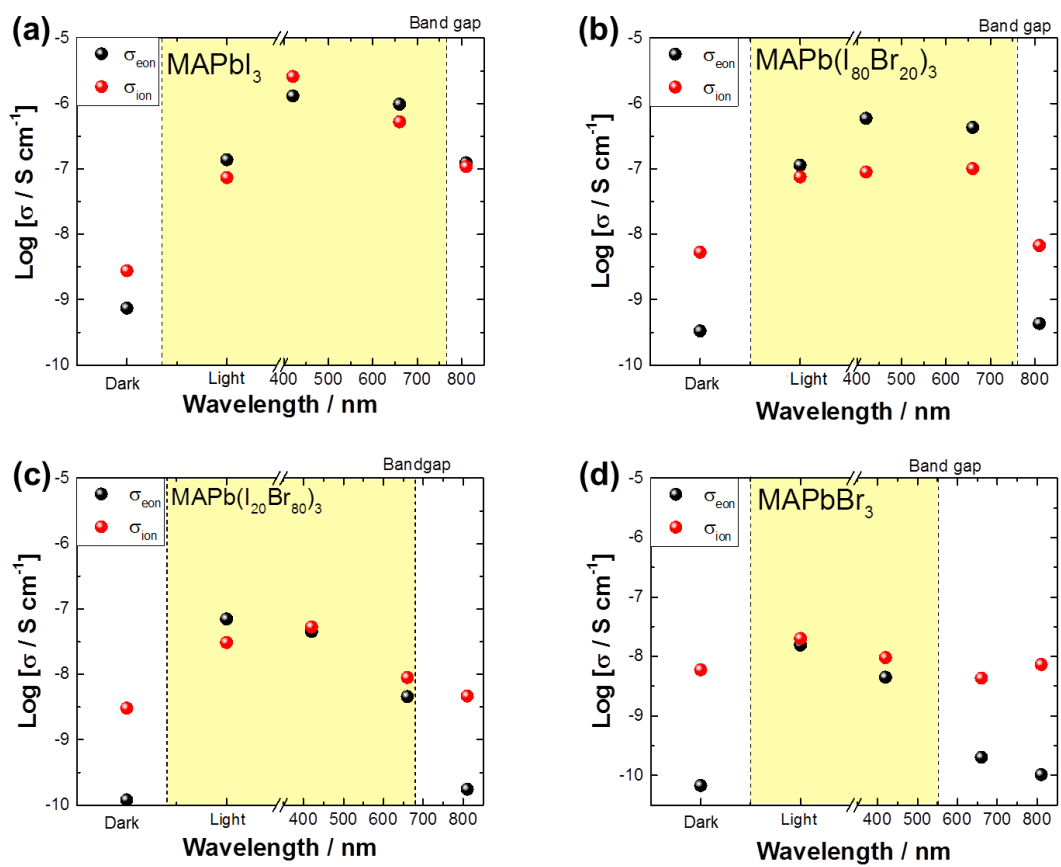

**Figure S5**. Wavelength dependence of electronic and ionic conductivities in MAPb(I,Br)$_3$. We used 420, 660, and 820 nm of LED lamp for light illumination with 1 mW/cm$^2$ of intensity. White light is also 1mW/cm$^2$ of light intensity. Yellow background indicates that photon energy is higher than bandgap. Ionic conductivity of MAPbBr$_3$ is not strongly enhanced under all wavelength of light.

## 5. Toluene experiments

To gain further evidence, we immersed $MAPbBr_3$ film into toluene solution, which acts as a solvent for $Br_2$. We measured the absorption signal of $Br_2$ in toluene as a function of time in the dark and under light. We can also compare the rate and quantify the bromine removal, which was already confirmed by $I_2$ in a previous work[6]. Unlike $MAPbI_3$, the amount of bromine removed from the samples is very weak under illumination. This result indicates a small light effect on bromide perovskite. In order to check the bromine signal in toluene, we dissolve $Br_2$ in toluene with 7 mmol and measure the absorption signal. As indicated in Fig. S6 inset, we can obtain clear absorption signal around 400 nm, which is in line with the literature data. Even though $Br_2$ has a high solubility in toluene[7], the excoporation rate of bromine under light is much weaker than iodine. In fact the bromine content in toluene was below our detection limit (see Fig. S6). This can be attributed to a very small homogeneity range of MAPbBr3 (*cf.* main text)

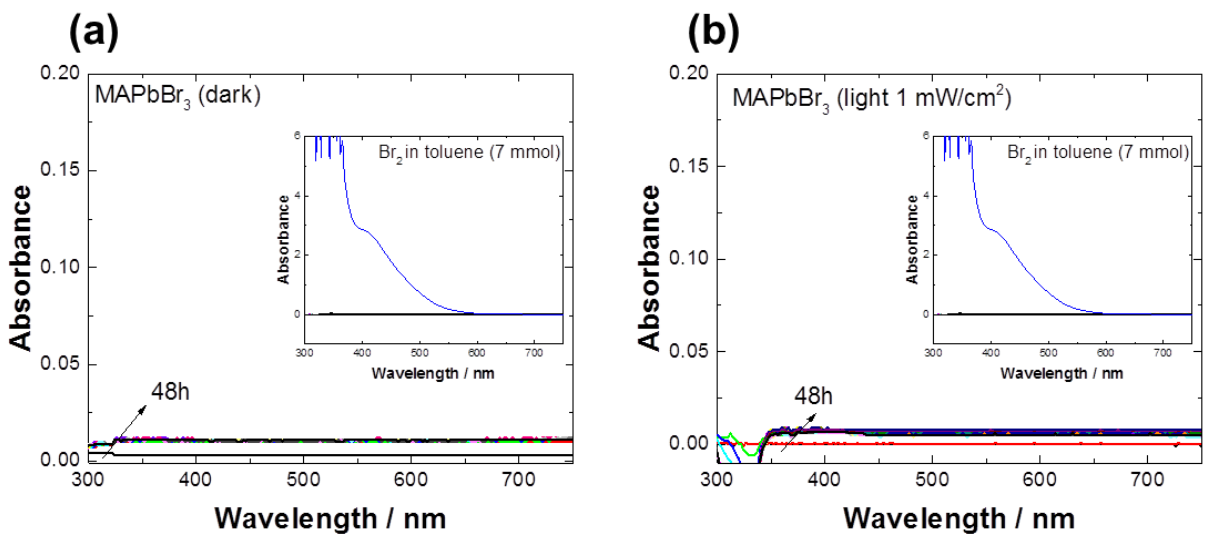

**Figure S6**. Absorbance of bromine from $MAPbBr_3$ thin films immersed in toluene (a) in the dark and (b) under light illumination. Standard bromine signal from $Br_2$ as shown in inset.

## 6. EMF of partially equilibration activity cell

We consider a mixed Br-conductor of thickness L (typically mm scale) in between to bromine partial pressures $P(Br_2)^/$ and $P(Br_2)^{//}$ (corresponding to the chemical potentials $\mu^{\cdot}$ and

$$\mu^{..} = \mu_{Br} = \mu^{\cdot} + \frac{1}{2}RT \ln P(Br_2)^{//}).$$

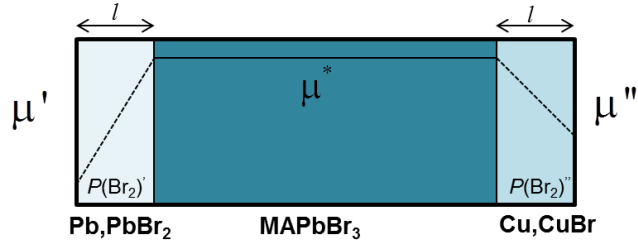

**Figure S7**. Schematic of *emf* measurement on MAPbBr$_3$ same as main text. We use Pb,PbBr$_2$ and Cu,CuBr pellets for the bromine partial pressures $P(Br_2)^/$ and $P(Br_2)^{//}$ and measure the open circuit voltage of this electrochemical cell. The partial pressure of each side is very low and the waiting time not sufficient to reach that equilibrium all through the sample so that we expect the real situation as shown $\mu^*$.

If the surface reaction is sufficiently fast and the waiting time for the whole sample to respond to the such conditions is sufficiently large ($\tau \gg L^2/D^\delta$), then

$$-\frac{EF}{RT} = \varepsilon = \int_{\mu^{\cdot}}^{\mu^{..}} t_{ion} d\mu \qquad (1).$$

Here $t_{ion}$ is the local transference number of bromine ions. It is convenient to convert Eq (1) as

$$\varepsilon = \bar{t}(\mu^{..} - \mu^{\cdot}) \qquad (2).$$

$\varepsilon$ is shown for -EF/RT; $\bar{t}$ is the mean ionic transference number. (Here, and in the following, we drop the index "ion" for simplicity) If however the waiting time is much less, such that equilibration from bulk sides only allows for equilibration within $l = \sqrt{2D^\delta t} < L$, then in the interior the remains a region determined by the initial partial pressure ($\mu^*$). This is realized in our case where $\mu^*$ corresponds to a high $P(Br_2)$ while $\mu'$, $\mu''$ correspond to very low $P(Br_2)$ values established by the redox couples (Pb:PbBr$_2$ or Cu:CuBr).

According to low Br-diffusivity of MAPbBr$_3$ the situation in the *emf* cell should be the one sketched in Fig. S7; very low Br-partial pressures (and very small values of t' and t'' owing to n-type conduction) on both sides, but high initial Br-partial pressure in the middle. What is the measured value? Eq(1) is still valid. Let us assume that μ increases linearly from the *l.h.s* value μ' (at x=0) to μ* (at x=l), stays constant there until x=L-l and then decreases to μ''. Without a qualitative restriction of quantity we assumed a linear behavior in the boundary regions, i.e.

$$\frac{dt}{d\mu} = const$$

. Accordingly, we decompose Eq. 2 into three regions

$$l.h.s \Rightarrow \varepsilon_l = \int_{\mu'}^{\mu^*} t_l \, d\mu$$

$$middle \Rightarrow \varepsilon_m = \int_{\mu^*}^{\mu^*} t^* \, d\mu = 0$$

$$r.h.s \Rightarrow \varepsilon_r = \int_{\mu^*}^{\mu''} t_r \, d\mu$$

The middle one drops out in the balance

$$t_l = \frac{t' - t^*}{\mu' - \mu^*} \mu + \frac{t^* \mu' - t' \mu^*}{\mu' - \mu^*} \quad (a_l \equiv \frac{t' - t^*}{\mu' - \mu^*}, b_l \equiv \frac{t^* \mu' - t' \mu^*}{\mu' - \mu^*})$$

$$t_r = \frac{t'' - t^*}{\mu'' - \mu^*} \mu + \frac{t^* \mu'' - t'' \mu^*}{\mu'' - \mu^*} \quad (a_r \equiv \frac{t'' - t^*}{\mu'' - \mu^*}, b_r \equiv \frac{t^* \mu'' - t'' \mu^*}{\mu'' - \mu^*})$$

According to Eq(1) the *emf* lowers as

$$\varepsilon = \int_{\mu'}^{\mu^*} (a_l \mu + b_l) d\mu + 0 + \int_{\mu^*}^{\mu''} (a_r \mu + b_r) d\mu = \frac{1}{2} a_l \mu^2 \Big]_{\mu'}^{\mu^*} + b_l \mu \Big]_{\mu'}^{\mu^*} + \frac{1}{2} a_r \mu^2 \Big]_{\mu^*}^{\mu''} + b_r \mu \Big]_{\mu^*}^{\mu''}$$

$$= \frac{1}{2} \frac{t' - t^*}{\mu' - \mu^*} (\mu^{*2} - \mu'^2) + \frac{t^* \mu' - t' \mu^*}{\mu' - \mu^*} (\mu^* - \mu') + \frac{1}{2} \frac{t'' - t^*}{\mu'' - \mu^*} (\mu^{*2} - \mu''^2) + \frac{t^* \mu'' - t'' \mu^*}{\mu'' - \mu^*} (\mu'' - \mu^*)$$

$$= \frac{1}{2} \frac{t' - t^*}{\mu' - \mu^*} (\mu^{*2} - \mu'^2) + \frac{1}{2} \frac{t'' - t^*}{\mu'' - \mu^*} (\mu^{*2} - \mu''^2) + \mu^* (t' - t'') + t^* (\mu'' - \mu')$$

Collecting terms we obtain

$$\varepsilon = \frac{1}{2} t^* (\mu'' - \mu') - \frac{1}{2} \mu^* (t'' - t') + \frac{1}{2} (t'' \mu'' - t' \mu')$$

If we introduce the condition $t^{'} \simeq t^{''} \simeq 0$ , the result is

$$\varepsilon = \frac{1}{2} t^{*} (\mu^{''} - \mu^{'})$$

Composition with Eq (2) shows that the effectively measured transference number is $\overline{t} = \frac{1}{2} t^{*}$

(we refer to this value as $\overline{t_{ion}}^{corr}$ in the main text) which satisfactorily agrees with the experiment results (see main text). The measured voltage value is hence much lower than the boundary values $t^{'}, t^{''}$ and less than the initial values.

A better approximation is obtained if only *t'* is set to zero corresponding to μ'< μ''. Then

$$\varepsilon = \frac{1}{2} t^{*} (\mu^{''} - \mu^{'}) - \frac{1}{2} \mu^{*} (t^{''} - t^{'}) + \frac{1}{2} (t^{'} \mu^{''} - t^{''} \mu^{'}) \text{ and hence } \overline{t} \leq \frac{1}{2} t^{*}.$$

## 7. Iodine partial pressure dependence and activation energy of FAPbI$_3$

We performed conductivity measurements under fixed iodine partial pressure (~$10^{-6}$ bar) on cation mixtures and measured activation energy of FAPbI$_3$. The ionic conductivity is reduced on increased FA concentration. The activation energy of FAPbI$_3$ is 0.6 eV, which is higher than for MAPbI$_3$ (0.4 eV). This reflects the high migration barrier for the I-vacancy motion owing to larger cation size.

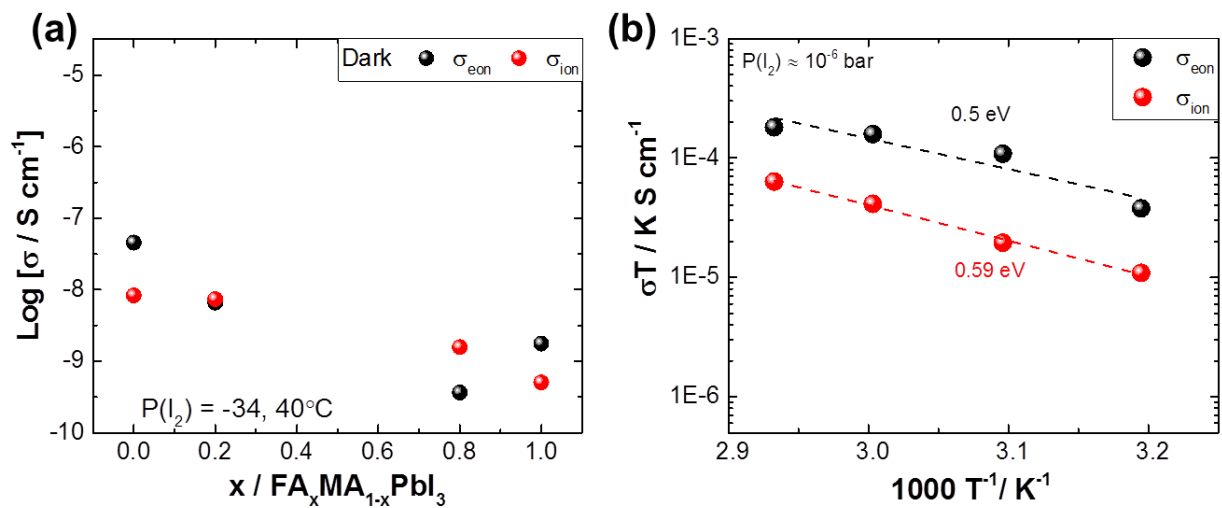

**Figure S8**. (a) Conductivity measurements as a function of iodine partial pressure and Ar as carrier gas for (MA,FA)PbI$_3$ thin films at 40°C. Ionic and electronic conductivity extracted from d.c. galvanostatic polarization. (b) Temperature dependence of electronic and ionic conductivities of FAPbI$_3$ in the dark under fixed iodine partial pressure from 40 to 70°C.

## 8. Polarization experiments in anion mixtures

To obtain a detailed explanation for photo de-mixing, we observed the dc-polarization conductivity variations followed by monitoring time dependence of halide mixtures under light. The polarization experiments give reliable $\tau^\delta$ and also $D^\delta$ values if x=0 or x=1. If however de-mixing occurs (x~0.5) then the relaxation times are increased by the morphological evolution. In that cases no true $D^\delta$-values can be extracted. But the unusually long $\tau^\delta$ values indicate the occurrence of de-mixing (see Fig. 5 in the main text).

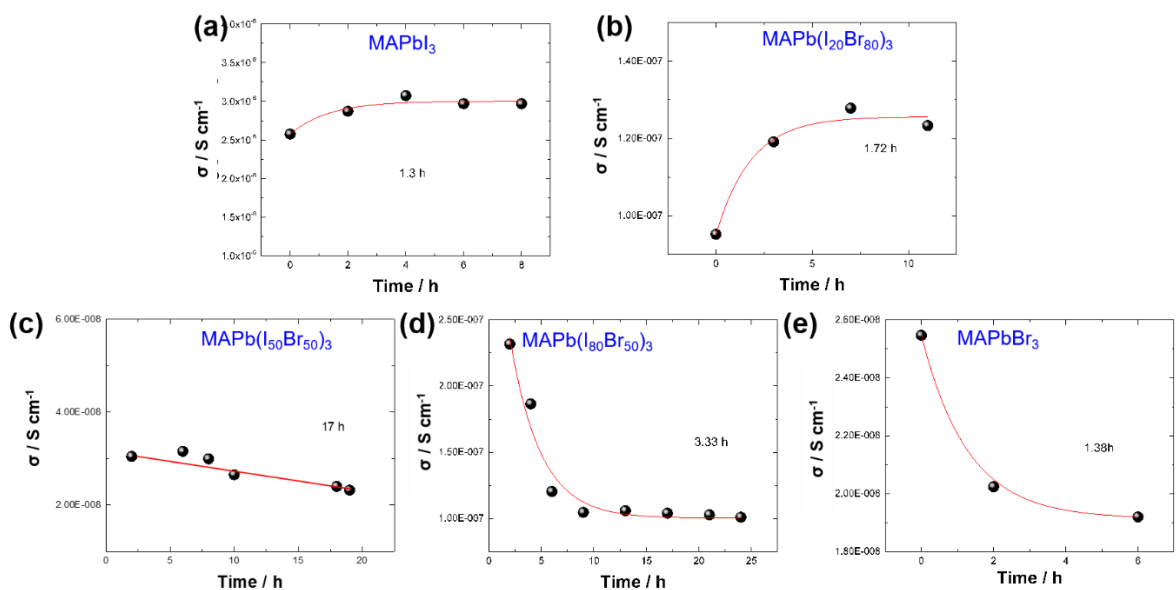

**Figure S9.** Equilibration of total conductivity of anion mixture perovskites under light illumination. We extracted the conductivity values from polarization measurement. Sample kept under Ar atmosphere with light illumination for ~20 hours during d.c. measurement. The similar equilibration time scale behavior upon illumination was observed in MAPbI$_3$ and MAPbBr$_3$. Longer equilibrium time scale was observed upon illumination in mixture sample ($x > 0.5$).

# 9. Self-trapped hole mechanism and photo de-mixing effect

Here we discuss a straight-forward explanation of photo-enhanced ion conductivity enabled by the hole trapping mechanism. The primary step of photoexcitation is the transfer of electrons from the valence to the conduction band which in chemical terms is approximately a charge transfer of electrons from the filled I-orbitals to empty Pb-orbitals (i). The next step is the localization of thus formed neutral iodine (ii) which owing to the smaller size can be accommodated interstitially. Herein further relaxation to $I_2^-$ dumbbells or higher aggregates $I_x^{(x-1)-}$ can occur (iii). In energetic terms we refer to (i) overcoming the band gap ($E_g$) (ii), losing half of the band width ($E_b$) (delocalization energy) and (iii) gaining the polarization energy ($E_r$). In alkali halides the last step is mainly attributed to $X_2^-$ molecule formation. Step (ii) involves the very narrow valence band as far as the holes are concerned, while for conduction electrons the wider width of the conduction band (together with the less favorable $E_r$) makes a self-trapping energetically unfavorable. For KI the total balance is favorable as far as hole self-trapping is concerned. Comparing KI with MAPI the major difference is the wider valence band of the perovskite. It is certainly a rough approximation to borrow the relaxation energy from KI, but this tentative approach already leads to a total self-trapping close to zero but slightly positive. A less favorable polarization energy for Br (lower polarizability, lower tendency to form higher aggregates) directly explains the much lower effects observed from $MAPbBr_3$. This simple mechanism also explains the insensitivity of the ionic conductivity effects if the cation (MA) is exchanged (FA). Moreover, the mechanism directly explains the observed reversible de-mixing of $MAP(I,Br)_3$ on illumination without resorting to elastic effects. Let us concentrate on the anions. Then the mixing process can, following Bragg and Williams, be viewed as a partial transformation of I-I and Br-Br bonds to I-Br bonds, the reaction energy ($2\Delta_R \varepsilon$)

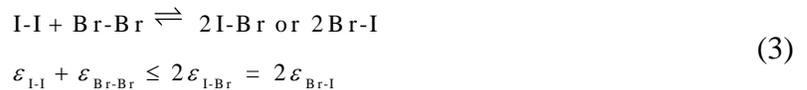

$$I\text{-}I + Br\text{-}Br \rightleftharpoons 2I\text{-}Br \text{ or } 2Br\text{-}I$$
$$\varepsilon_{I\text{-}I} + \varepsilon_{Br\text{-}Br} \leq 2\varepsilon_{I\text{-}Br} = 2\varepsilon_{Br\text{-}I}$$

(3)

describing the interaction between the two anions. Approximating the configurational term in the mixing free energy $F_m$ by using ideal mixing entropy and assuming a positive $\Delta_R \varepsilon$ (repulsion), $F_m(x_{Br})$ has the double minimum form. The half of the reaction energy, i.e. $\Delta_R\varepsilon$, determines the critical temperature below which de-mixing occurs $(T_c \propto \Delta_R\varepsilon)$. In the dark $\Delta_R\varepsilon$ is obviously small such that at room temperature all mixtures are stable ($T_c <$ room temperature).

On illumination, Eq. (3) refers to the excited states. The I-I pair is stabilized by the self-trapping energy weighted by the probability of forming a hole (hole concentration per I-concentration), while the effect on the Br-Br pair may be neglected. Moreover a replacement of I in the I-I pair by a Br will destroy the energetic effect. Hence $\Delta_R\varepsilon$ measures the loss of the energetic stabilization $\left(\delta\Delta_R\varepsilon \cong \Delta_R\varepsilon \cong -\frac{1}{2}w\right)$ where $\delta\Delta_R\varepsilon$ is excess reaction energy under light and $w$ regular solution interaction energy parameter of the I-I and Br-Br pairs.

In order to bring $T_c$ above room temperature, a $w$-value on the order of 10 meV is necessary which is consistent with the above considerations. It is noteworthy that this effect also explains the disappearance of the de-mixing at grain sizes below 50 nm. The simple reason is that in the above analysis the cost of the phase boundary formed on de-mixing was neglected. For a 50 nm grain the interfacial energy is $\dfrac{G^\Sigma}{V/V_m} \sim \dfrac{\gamma}{L}V_m$ ($G^\Sigma$: Excess Gibbs free energy per unit area, V$_m$: molar volume, L: grain size, $\gamma$: interfacial tension). Using typical values for $\gamma$ one obtains indeed values of the order of $w$. For the same reason, the segregation into I-rich and Br-rich phases will not stop at small cluster sizes but develop into domains greater than 50 nm. Unfortunately more precise values are not available for checking if the described effect suffices or has to be implemented by elastic effects in order to qualitatively explain the features. A question that remains is if also e$^-$ is self-trapped in the photo-perovskites. Unlike the alkali-halides the differences in the band widths of valence and conduction band are not that different. The possibly lower ability of Pb$^+$ or Pb$^0$ to be stabilized by additional lead may be compensated by a stabilization through the $I_i^o$.

In the extreme case that both electrons and holes are strongly self-trapped the photo-excitation would then be

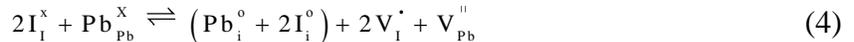

$$2I_I^x + Pb_{Pb}^X \rightleftharpoons \left(Pb_i^o + 2I_i^o\right) + 2V_I^\cdot + V_{Pb}^{''} \tag{4}$$

Unlike $v_I^{\cdot}$, $v_{Pb}^{\prime\prime}$ is comparatively immobile. Interestingly Eq. (4) is a photoexcitation reaction in which $e^{\prime}$ and $h^{\cdot}$ do not explicitly occur; this would be in agreement with the large enhancement of the ionic defect concentration beyond the remaining electronic level.

Modelling by Ginsberg[8,9] et al invokes strain effects coupled to the electronic excitation and polarization. Photo-striction effects have been observed on illumination, and it may in fact be tempting to correlate it with our mechanism of self-trapping. It is well established that illumination increases the volume of the iodide by about 0.02 % and the effect on the bromide is a bit less than iodide[10-12]. Light illumination leads to either less-distorted Pb–I–Pb bonds or extension of the Pb–I bonds, which induce lattice expansion. Here it is revealing to refer to Darken's model[13] of de-mixing of a model if oversized substitutions are introduced. This leads to an overall volume effect that is characterized by the Poisson ratio of the matrix. The resulting mixture is approximated by a regular solution model characterized by an interaction parameter v (Poisson's ratio). A negligible v leads to a de-mixing at $\Delta V_2 / \Delta V_1 = 3(1-v)/(1+v)$. If we identify $\Delta V_2 / \Delta V_1$ with room temperature and associate $\Delta V_2 / \Delta V_1$ with the volume effect, Darken has established for a size mismatch of 5% solubility limit of $\sim 15\%$. In our case the quantitative situation is certainly different, but the qualitative effect is the same if we replace the substitution by the illumination effect. In order for the small strain effects observed in the perovskites to account for the de-mixing, the Gibbs free energy of mixing must be very small.

Certainly, the combination of strain and chemical effects is to be considered for a refined model of photo-demixing[14].

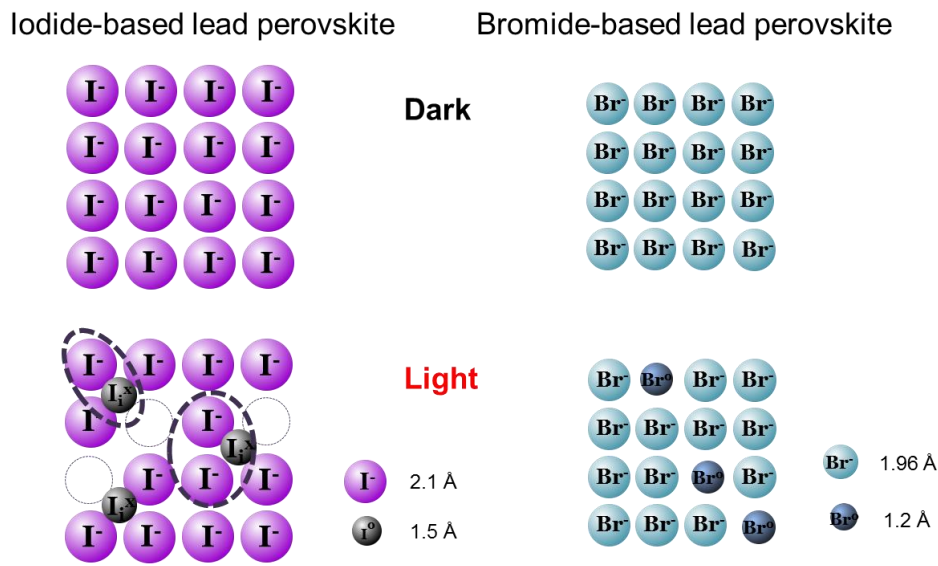

**Figure S10.** Schematic of self-trapping hole mechanism of iodide and bromide perovskites under dark and light conditions.